\documentclass[aps,rpl,preprint,groupedaddress]{revtex4}
\begin{document}

\title{Quantum teleportation via two qubit Heisenberg $XY$ chain - Effects of anisotropy and magnetic field}
\author{Ye Yeo}
\affiliation{Centre for Mathematical Sciences, Wilberforce Road, Cambridge CB3 0WB, United Kingdom and \\
Department of Physics, National University of Singapore, 10 Kent Ridge Crescent, Singapore 119260, Singapore}

\author{Tongqi Liu}
\affiliation{Department of Engineering, Trumpington Street, Cambridge CB3 1PZ, United Kingdom}

\author{Yu-En Lu}
\affiliation{Computer Laboratory, William Gates Building, 15 J J Thomson Avenue, Cambridge CB3 0FD, United Kingdom}

\author{Qi-Zhong Yang}
\affiliation{Cavendish Laboratory, Madingley Road, Cambridge CB3 0HE, United Kingdom}

\begin{abstract}
In this paper we study the influence of anisotropy on the usefulness, of the entanglement in a two-qubit Heisenberg $XY$ chain at thermal equilibrium in the presence of an external magnetic field, as resource for quantum teleportation via the standard teleportation protocol.  We show that the nonzero thermal entanglement produced by adjusting the external magnetic field beyond some critical strength is a useful resource.  We also considered entanglement teleportation via two two-qubit Heisenberg $XY$ chains.
\end{abstract}

\maketitle

\section{Introduction}
The one-dimensional Heisenberg models have been extensively studied in solid state physics (see references in \cite{Arnesen}).  Interest in these models has been revived recently by several proposals for realizing quantum computation \cite{Loss} and information processing \cite{Imamoglu} using quantum dots (localized electron spins) as qubits \cite{Schumacher}.  Lying at the heart of quantum computation and quantum information \cite{Nielsen} is a physical resource - quantum entanglement.  Consequently, entanglement in interacting Heisenberg spin systems at finite temperatures has been investigated by a number of authors (see, e.g., Ref.\cite{Osborne} and references therein).

The state of a typical solid state system at thermal equilibrium (temperature $T$) is $\chi = e^{-\beta H}/Z$, where $H$ is the Hamiltonian, $Z = {\rm tr}e^{-\beta H}$ is the partition function, and $\beta = 1/kT$, where $k$ is the Boltzmann's constant.  The entanglement associated with the thermal state $\chi$ is referred to as the thermal entanglement \cite{Arnesen}.  Due to the availability of a good and computable measure of entanglement for systems of two qubits, the concurrence (see Section II), thermal entanglement in two-qubit Heisenberg spin chains has been thoroughly analyzed in terms of thermal concurrence \cite{Arnesen, Wang1, Wang2, Gunlycke, Kamta, Anteneodo, Zhou}.  In particular, Kamta {\em et al.} \cite{Kamta} showed that the anisotropy and the magnetic field strength may together be used to control the extent of thermal concurrence in a two-qubit Heisenberg $XY$ chain and, especially, to produce entanglement for any finite $T$, by adjusting the external magnetic field beyond some critical strength.  The recent breakthroughs in the experimental physics of double quantum dot (see, e.g., Ref.\cite{Chen}) shows that these studies are worthwhile pursuits.

An entangled composite system gives rise to nonlocal correlation between its subsystems that does not exist classically.  This nonlocal property enables the uses of local quantum operations and classical communication to teleport an unknown quantum state via a shared pair of entangled particles with fidelity (see Section IV) better than any classical communication protocol \cite{Bennett, Popescu, Horodecki}.  Quantum teleportation can thus serve as an operational test of the presence and ``quality'' of entanglement.  On the other hand, two-qubit teleportation together with one-qubit unitary operations are sufficient to implement the universal gates for quantum computation \cite{Gottesman}.  Hence, from both fundamental and practical viewpoints, it is important to study the thermally mixed entangled state of a Heisenberg spin system via quantum teleportation.  The possibility of using the thermally mixed entangled state of a two-qubit Heisenberg $XX$ chain as resource for the standard teleportation protocol ${\cal P}_0$ \cite{Bennett} was considered in Ref.\cite{Yeo1}.  The local quantum operations in ${\cal P}_0$ consist of Bell measurements and Pauli rotations.  It was shown that although quantum teleportation with fidelity better than any classical communication protocol is possible, the amount of nonzero thermal entanglement does not guarantee this.  We could have a more entangled thermal state not achieving a better fidelity than a less entangled one.  In fact, the thermally mixed entangled state of a two-qubit Heisenberg $XX$ chain is ``useless'' whenever an external magnetic field above some critical strength is applied \cite{Yeo1}.  Entanglement teleportation \cite{Lee} using the thermally mixed entangled states of two two-qubit Heisenberg $XX$ chains as resources was also studied in Ref.\cite{Yeo2}.

In view of the above results, we study in this paper the influence of anisotropy on the usefulness, of the entanglement in a two-qubit Heisenberg $XY$ chain at thermal equilibrium in the presence of an external magnetic field, as resource for quantum teleportation via the standard teleportation protocol ${\cal P}_0$.  This paper is organized as follows.  In Section II, we give the Hamiltonian for the anisotropic two-qubit Heisenberg $XY$ chain, and briefly review a measure of entanglement, the concurrence.  We briefly discuss the relevant results of Ref.\cite{Kamta} in Section III.  This sets the stage necessary for the presentation of our results in Section IV.  In Section V, we present our results on entanglement teleportation using the thermally mixed entangled states of two two-qubit Heisenberg $XY$ chains as resources.  We conclude in Section VI.

\section{Two qubit Heisenberg $XY$ chain}
The Hamiltonian $H$ for the anisotropic two-qubit Heisenberg $XY$ chain in an external magnetic field $B_m \equiv \eta J$ ($\eta$ is a real number) along the $z$ axis is
\begin{equation}
H = \frac{1}{2}(1 + \gamma)J\sigma^1_A \otimes \sigma^1_B + \frac{1}{2}(1 - \gamma)J\sigma^2_A \otimes \sigma^2_B + \frac{1}{2}B_m(\sigma^3_A \otimes \sigma^0_B + \sigma^0_A \otimes \sigma^3_B),
\end{equation}
where $\sigma^0_{\alpha}$ is the identity matrix and $\sigma^i_{\alpha}$ $(i = 1, 2, 3)$ are the Pauli matrices at site $\alpha = A, B$.  The parameter $-1 \leq \gamma \leq 1$ measures the anisotropy of the system and equals $0$ for the isotropic $XX$ model \cite{Wang1} and $\pm 1$ for the Ising model \cite{Gunlycke}.  $(1 + \gamma)J$ and $(1 - \gamma)J$ are real coupling constants for the spin interaction.  The chain is said to be antiferromagnetic for $J > 0$ and ferromagnetic for $J < 0$.

The eigenvalues and eigenvectors of $H$ are given by \cite{Kamta}
\begin{eqnarray}
H|\Phi^0\rangle_{AB} & = & {\cal B} |\Phi^0\rangle_{AB}, \nonumber \\
H|\Phi^1\rangle_{AB} & = & J        |\Phi^1\rangle_{AB},\ \nonumber \\
H|\Phi^2\rangle_{AB} & = & -J       |\Phi^2\rangle_{AB},\ \nonumber \\
H|\Phi^3\rangle_{AB} & = & -{\cal B}|\Phi^3\rangle_{AB},
\end{eqnarray}
where ${\cal B} \equiv \sqrt{B^2_m + \gamma^2J^2} = \sqrt{\eta^2 + \gamma^2}J$,
\begin{eqnarray}
|\Phi^0\rangle_{AB} & = & \frac{1}{\sqrt{({\cal B} + B_m)^2 + \gamma^2J^2}}
[({\cal B} + B_m)|00\rangle_{AB} + \gamma J|11\rangle_{AB}], \\
|\Phi^1\rangle_{AB} & = & \frac{1}{\sqrt{2}}[|01\rangle_{AB} + |10\rangle_{AB}], \\
|\Phi^2\rangle_{AB} & = & \frac{1}{\sqrt{2}}[|01\rangle_{AB} - |10\rangle_{AB}], \\
|\Phi^3\rangle_{AB} & = & \frac{1}{\sqrt{({\cal B} - B_m)^2 + \gamma^2J^2}}
[({\cal B} - B_m)|00\rangle_{AB} - \gamma J|11\rangle_{AB}].
\end{eqnarray}
When $B_m = 0$, Eqs.(3) and (6) reduce to $\frac{1}{\sqrt{2}}[|00\rangle_{AB} + |11\rangle_{AB}]$ and $\frac{1}{\sqrt{2}}[|00\rangle_{AB} - |11\rangle_{AB}]$ respectively, so that the eigenvectors are the four maximally entangled Bell states: $|\Psi^0_{Bell}\rangle_{AB}$, $|\Psi^1_{Bell}\rangle_{AB}$, $|\Psi^2_{Bell}\rangle_{AB}$, and $|\Psi^3_{Bell}\rangle_{AB}$.  For $\gamma = 0$, $|\Phi^0\rangle_{AB} = |00\rangle_{AB}$ and $|\Phi^3\rangle_{AB} = |11\rangle_{AB}$ with eigenvalues $B_m$ and $-B_m$ respectively, while $|\Phi^1\rangle_{AB}$ and $|\Phi^2\rangle_{AB}$ remain unchanged \cite{Wang1, Yeo1}.

To quantify the amount of entanglement associated with a given two-qubit state $\rho_{AB}$, we consider the concurrence \cite{Hill, Wootters} ${\cal C}[\rho_{AB}] \equiv \max\{\lambda_1 - \lambda_2 - \lambda_3 - \lambda_4,\ 0\}$, where $\lambda_k$ $(k = 1, 2, 3, 4)$ are the square roots of the eigenvalues in decreasing order of magnitude of the spin-flipped density-matrix operator $R_{AB} = \rho_{AB}(\sigma^2_A \otimes \sigma^2_B)\rho^*_{AB}(\sigma^2_A \otimes \sigma^2_B)$, the asterisk indicates complex conjugation.  The concurrence associated with the eigenvectors, Eqs.(3) and (6), are given by $\frac{\gamma}{\sqrt{\eta^2 + \gamma^2}}$.  Hence, they represent entangled states when $\gamma \not= 0$.  We note that in the limit of large $\eta$,
\begin{equation}
{\cal C}[|\Phi^0\rangle_{AB}\langle\Phi^0|] = {\cal C}[|\Phi^3\rangle_{AB}\langle\Phi^3|] \approx \gamma\eta^{-1},
\end{equation}
going to zero asympotically when $\eta$ is infinitely large.

\section{Thermal state and concurrence}
For the above system in thermal equilibrium at temperature $T$, its state is described by the density operator
\begin{equation}
\chi_{AB} = \frac{1}{Z}[e^{-\beta{\cal B}}|\Phi^0\rangle_{AB}\langle\Phi^0| + e^{-\beta J}|\Phi^1\rangle_{AB}\langle\Phi^1| + e^{\beta J}|\Phi^2\rangle_{AB}\langle\Phi^2| + e^{\beta{\cal B}}|\Phi^3\rangle_{AB}\langle\Phi^3|],
\end{equation}
where the partition function $Z = 2\cosh\beta{\cal B} + 2\cosh\beta J$, the Boltzmann's constant $k \equiv 1$ from hereon, and $\beta = 1/T$.  After some straightforward algebra, we obtain
\begin{eqnarray}
\lambda_1 & = & \frac{1}{Z}e^{\beta J}, \\
\lambda_2 & = & \frac{1}{Z}e^{-\beta J}, \\
\lambda_3 & = & \frac{1}{Z}\sqrt{1 + \frac{2\gamma^2J^2}{{\cal B}^2}\sinh^2\beta{\cal B} + \frac{2\gamma J}{{\cal B}}\sqrt{1 + \frac{\gamma^2J^2}{{\cal B}^2}\sinh^2\beta{\cal B}}\sinh\beta{\cal B}}, \\
\lambda_4 & = & \frac{1}{Z}\sqrt{1 + \frac{2\gamma^2J^2}{{\cal B}^2}\sinh^2\beta{\cal B} - \frac{2\gamma J}{{\cal B}}\sqrt{1 + \frac{\gamma^2J^2}{{\cal B}^2}\sinh^2\beta{\cal B}}\sinh\beta{\cal B}}.
\end{eqnarray}
The concurrence derived from Eqs.(9) - (12) is invariant under the substitutions $\eta \longrightarrow -\eta$, $\gamma \longrightarrow -\gamma$, and $J \longrightarrow -J$.  We shall see that the same applies to the fully entangled fraction (see Section IV).  Therefore, we restrict our considerations to $\eta \geq 0$, $0 \leq \gamma \leq 1$, and $J > 0$.

For $\gamma = 0$, Eqs.(9) - (12) yield the thermal concurrence obtained in Ref.\cite{Wang1}:
\begin{equation}
{\cal C}[\chi_{AB}] = \left\{\frac{\sinh\beta J - 1}{\cosh\beta B_m + \cosh\beta J},\ 0\right\}.
\end{equation}
Eq.(13) reduces, when $T = 0$, to
\begin{equation}
{\cal C}[\chi_{AB}] = \left\{\begin{array}{lll}
1           & {\rm for} & \eta < 1, \\
\frac{1}{2} & {\rm for} & \eta = 1, \\
0           & {\rm for} & \eta > 1.
\end{array}\right.
\end{equation}
Hence, with increasing $\eta$, the concurrence is constant and maximal, but it drops suddenly to zero as $\eta$ crosses the critical value $\eta_{\rm critical} = 1$, marking the point of quantum phase transition (phase transition taking place at zero temperature due to variation of interaction terms in the Hamiltonian of a system \cite{Arnesen}).  For nonzero $T$ and, $\eta < 1$ or $\eta = 1$, ${\cal C}[\chi_{AB}]$ decreases from $1$ or $\frac{1}{2}$ respectively to $0$ as $T$ increases from $0$ to the critical temperature $T^{(1)}_{\rm critical} \approx 1.13459J$.  Physically, this is due to mixing of unentangled states and entangled ones.  Similarly, when $\eta > 1$, due to mixing of entangled states with the original product state $|11\rangle_{AB}$, ${\cal C}[\chi_{AB}]$ increases from $0$ to some maximum before decreasing to $0$ again at the same $T^{(1)}_{\rm critical}$.  The critical temperature $T^{(1)}_{\rm critical}$, beyond which ${\cal C}[\chi_{AB}] = 0$, is independent of $B_m$.  It also follows from Eq.(13) that ${\cal C}[\chi_{AB}]$ decreases monotonically with increasing $B_m$ for any finite $T$ and vanishes exponentially with increasing $B_m$, due to an increase in the proportion of $|11\rangle_{AB}$.

For some nonzero $\gamma$, Eq.(8) reduces to the following three possibilities in the zero-temperature limit, i.e., $\beta \longrightarrow \infty$, at which the system is in its ground state.
\begin{description}
\item{(a)} $0 \leq \eta < \sqrt{1 - \gamma^2}$:
\begin{eqnarray}
\chi_{AB} & = & \frac{1}{Z}[e^{\beta J}|\Phi^2\rangle_{AB}\langle\Phi^2| + e^{\beta{\cal B}}|\Phi^3\rangle_{AB}\langle\Phi^3|] 
\nonumber \\
& \longrightarrow & |\Phi^2\rangle_{AB}\langle\Phi^2|,
\end{eqnarray}
with $Z = e^{\beta J} + e^{\beta\cal B}$.  Eqs.(9) - (12) give ${\cal C}[\chi_{AB}] = 1$, its maximum value, in agreement with the fact that $|\Phi^2\rangle_{AB}$ is a maximally entangled Bell state.
\item{(b)} $\eta = \sqrt{1 - \gamma^2}$:
\begin{equation}
\chi_{AB} \longrightarrow \frac{1}{2}[|\Phi^2\rangle_{AB}\langle\Phi^2| + |\Phi^3\rangle_{AB}\langle\Phi^3|].
\end{equation}
From Eqs.(9) - (12), the above equally weighted mixture has
\begin{equation}
{\cal C}[\chi_{AB}] = \frac{1}{2}(1 - \gamma).
\end{equation}
\item{(c)} $\eta > \sqrt{1 - \gamma^2}$:
\begin{equation}
\chi_{AB} \longrightarrow |\Phi^3\rangle_{AB}\langle\Phi^3|,
\end{equation}
and Eqs.(9) - (12) yield accordingly
\begin{equation}
{\cal C}[\chi_{AB}] = \frac{\gamma}{\sqrt{\eta^2 + \gamma^2}}.
\end{equation}
\end{description}
Therefore, for a given $\gamma$, $\eta_{\rm critical} = \sqrt{1 - \gamma^2}$ marks the point of quantum phase transition.  However, in contrast to the isotropic case ($\gamma = 0$) \cite{Wang1}, this is not a transition from an entangled phase to an unentangled phase.  For values of $\gamma$ other than $\gamma = \frac{1}{3}$, there is a sudden increase or decrease in ${\cal C}[\chi_{AB}]$ at $\eta_{\rm critical}$, depending on whether $\gamma > \frac{1}{3}$ or $\gamma < \frac{1}{3}$, before decreasing to zero asymptotically, as $\eta$ is increased beyond the critical value $\eta_{\rm critical}$ \cite{Kamta}.

For nonzero temperatures, due to mixing, ${\cal C}[\chi_{AB}]$ decreases to zero as the temperature $T$ is increased beyond the critical value $T^{(1)}_{\rm critical}$, as in the isotropic case.  However, in contrast, for each nonzero value of $\gamma$, $T^{(1)}_{\rm critical}$ depends on $\eta$.  It decreases with $\eta$ when $\eta$ is increased from $0$ to $\eta_{\rm critical} = \sqrt{1 - \gamma^2}$, but increases with $\eta$ for $\eta > \eta_{\rm critical}$.  In fact, the anisotropy permits one to obtain entangled qubits at higher $T$ and higher $B_m$ than is possible in the isotropic case \cite{Kamta}.  An interesting question therefore is whether this entanglement is ``useful'' as a resource for teleportation.  This is the subject of our next two sections.

\section{Teleportation and fully entangled fraction}
Standard teleportation ${\cal P}_0$ \cite{Bennett} with an arbitrary entangled mixed state resource $\rho_{AB}$ is equivalent to a generalized depolarizing channel $\Lambda^{\rho_{AB},\ {\cal P}_0}_B$, with probabilities given by the maximally entangled components of the resource \cite{Bowen, Albeverio}.  Therefore, for the thermally mixed entangled state $\chi_{AB}$, Eq.(8), we have
\begin{equation}
\Lambda^{\chi_{AB},\ {\cal P}_0(m)}_B(|\psi\rangle_B\langle\psi|) = \sum^3_{i = 0}
{_{AB}}\langle\Psi^i_{Bell}|\chi_{AB}|\Psi^i_{Bell}\rangle_{AB} \times
\sigma^{i\oplus m}_B|\psi\rangle_B\langle\psi|\sigma^{i\oplus m}_B,
\end{equation}
where $|\psi\rangle_B = \cos\frac{\vartheta}{2}|0\rangle_B + e^{i\varphi}\sin\frac{\vartheta}{2}|1\rangle_B$ ($0 \leq \vartheta \leq \pi$, $0 \leq \varphi \leq 2\pi$) is an arbitrary unknown pure state of a qubit.  Here, $i \oplus m$ denotes summation modulus $4$, with $m = 0, 1, 2, 3$.  In this paper, reliability for teleportation will be the criterion for judging the quality of the entangled thermal state, Eq.(8).  Quantitatively, this is measured by the teleportation fidelity,
\begin{equation}
\Phi[\Lambda^{\chi_{AB},\ {\cal P}_0(m)}_B] \equiv \int d\psi\ {_B}\langle\psi|\Lambda^{\chi_{AB},\ {\cal P}_0(m)}_B(|\psi\rangle_B\langle\psi|)|\psi\rangle_B.
\end{equation}
In the standard teleportation protocol ${\cal P}_0$, the maximal teleportation fidelity $\Phi_{\max}[\Lambda^{\chi_{AB},\ {\cal P}_0}_B]$ achievable is given by \cite{Horodecki, Albeverio}
\begin{equation}
\Phi_{\max}[\Lambda^{\chi_{AB},\ {\cal P}_0}_B] = \frac{2{\cal F}[\chi_{AB}] + 1}{3},
\end{equation}
where the fully entangled fraction
\begin{equation}
{\cal F}[\chi_{AB}] \equiv \max_{i = 0, 1, 2, 3}\{{_{AB}}\langle\Psi^i_{Bell}|\chi_{AB}|\Psi^i_{Bell}\rangle_{AB}\}.
\end{equation}
After some straightforward algebra, we obtain
\begin{equation}
{\cal F}[\chi_{AB}] = \max\left\{\frac{1}{Z}e^{\beta J},\ \frac{1}{Z}\left(\cosh\beta{\cal B} + \frac{\gamma J}{\cal B}\sinh\beta{\cal B}\right),\ \frac{1}{Z}e^{-\beta J},\ \frac{1}{Z}\left(\cosh\beta{\cal B} - \frac{\gamma J}{\cal B}\sinh\beta{\cal B}\right)\right\}.
\end{equation}
So, ${\cal F}[\chi_{AB}]$ is indeed invariant under the substitutions $\eta \longrightarrow -\eta$, $\gamma \longrightarrow -\gamma$, and $J \longrightarrow -J$.  Restricting our considerations to $\eta \geq 0$, $0 \leq \gamma \leq 1$, and $J > 0$, we have
\begin{equation}
{\cal F}[\chi_{AB}] = \left\{\begin{array}{lll}
\frac{1}{Z}e^{\beta J}                                                                & {\rm for} & \sqrt{\eta^2 + \gamma^2} \leq 1, \\
\frac{1}{Z}\left(\cosh\beta{\cal B} + \frac{\gamma J}{\cal B}\sinh\beta{\cal B}\right)& {\rm for} & \sqrt{\eta^2 + \gamma^2} > 1.
\end{array}\right.
\end{equation}
For a bipartite entangled state $\rho_{AB}$ to be useful for quantum teleportation we must have the fully entangled fraction ${\cal F}[\rho_{AB}] > \frac{1}{2}$ \cite{Popescu, Horodecki}.

In the zero temperature limit, Eq.(25) reduces to
\begin{equation}
{\cal F}[\chi_{AB}] = \left\{\begin{array}{lll}
1                                                                                 & {\rm for} & \sqrt{\eta^2 + \gamma^2} < 1, \\
\frac{1}{2}                                                                       & {\rm for} & \sqrt{\eta^2 + \gamma^2} = 1, \\
\frac{1}{2}\left(1 + \frac{\gamma}{\sqrt{\eta^2 + \gamma^2}}\right) > \frac{1}{2} & {\rm for} & \sqrt{\eta^2 + \gamma^2} > 1.
\end{array}\right.
\end{equation}
Therefore, the equally weighted mixture Eq.(16) is useless for quantum teleportation even though its concurrence Eq.(17) may not be zero.  For a given value of $\gamma$, $\eta_{\rm critical} = \sqrt{1 - \gamma^2}$ again marks the point of quantum phase transition from the maximally entangled state Eq.(15), which yields ${\cal F}[\chi_{AB}] = 1$, to the generally nonmaximally entangled state Eq.(18), which still yields ${\cal F}[\chi_{AB}] > \frac{1}{2}$ as long as $\eta$ is finite.  In the limit of large $\eta$,
\begin{equation}
{\cal F}[\chi_{AB}] \approx \frac{1}{2} + \frac{1}{2}\gamma\eta^{-1} > \frac{1}{2},
\end{equation}
equals $\frac{1}{2}$ asymptotically when $\eta$ is infinitely large.  This is in contrast to the isotropic case, where the quantum phase transition occurs at $\eta_{\rm critical} = 1$, from the maximally entangled phase to the unentangled phase (see Eq.(14)).  Obviously, in this case, ${\cal F}[\chi_{AB}]$ jumps from $1$ to $\frac{1}{2}$ at $\eta_{\rm critical} = 1$.  And ${\cal F}[\chi_{AB}] = \frac{1}{2}$ when $\eta$ is increased beyond $1$ \cite{Yeo1}.

At nonzero temperatures, due to mixing, ${\cal F}[\chi_{AB}]$ decreases to $\frac{1}{2}$ at the critical temperature $T^{(2)}_{\rm critical}$ beyond which the performance is worse than what classical communication protocol can offer, as in the isotropic case \cite{Yeo1}.  Also, for each $\gamma$, $T^{(2)}_{\rm critical}$ is dependent on $\eta$.  To obtain $T^{(2)}_{\rm critical}$ we consider the following.  For the thermal state Eq.(8) to be useful for quantum teleportation at nonzero $T$, when $\sqrt{\eta^2 + \gamma^2} < 1$,  we demand that
\begin{equation}
\sinh\beta J > \cosh\beta\sqrt{\eta^2 + \gamma^2}J,
\end{equation}
and  when $\sqrt{\eta^2 + \gamma^2} > 1$, we demand that
\begin{equation}
\frac{\gamma}{\sqrt{\eta^2 + \gamma^2}}\sinh\beta\sqrt{\eta^2 + \gamma^2}J > \cosh\beta J.
\end{equation}
When $\gamma = 0$, Eq.(28) can be satisfied as long as $\eta < 1$, but Eq.(29) is unattainable.  $T^{(2)}_{\rm critical}$ decreases from $1.13459J$ to $0$ as $\eta$ is increased from $0$ to $\eta_{\rm critical} = 1$, and remains zero when $\eta$ is increased beyond $1$ \cite{Yeo1}.  For nonzero $\gamma$, $T^{(2)}_{\rm critical}$ similarly decreases to zero when $\eta$ is increased from $0$ to $\eta_{\rm critical} = \sqrt{1 - \gamma^2}$, as shown in Figure 1.  However, this is followed by a monotonic increase in $T^{(2)}_{\rm critical}$ as $\eta$ is increased beyond $\sqrt{1 - \gamma^2}$, in contrast to the isotropic case.  The behavior of $T^{(2)}_{\rm critical}$ is therefore qualitatively similar to $T^{(1)}_{\rm critical}$.  In the limit of large $\eta$,
\begin{equation}
T^{(2)}_{\rm critical} \approx \frac{\eta J}{\ln\eta - \ln\gamma + \ln2}.
\end{equation}
Therefore, the anisotropy not only allows one to obtain entangled qubits at higher $T$ and higher $B_m$ than is possible in the isotropic case, but also the associated entanglement is useful as resource for teleportation via ${\cal P}_0$.

\section{Entanglement teleportation}
Lee and Kim \cite{Lee} considered teleportation of an entangled two-body pure spin-$\frac{1}{2}$ state via two independent, equally entangled, noisy quantum channels represented by Werner states \cite{Werner}.  In their two-qubit teleportation protocol ${\cal P}_1$, the joint measurement is decomposable into two independent Bell measurements and the unitary operation into two local one-qubit Pauli rotations.  In other words, ${\cal P}_1$ is a straightforward generalization of the standard teleportation protocol ${\cal P}_0$, just doubling the setup.  However, they found that quantum entanglement of the two-qubit state is lost during the teleportation even when the channel has nonzero quantum entanglement, and in order to teleport quantum entanglement the quantum channels should possess a critical value of minimum entanglement.  Hence, teleportating entanglement demands more stringent conditions on the quantum channels.  Mathematically, we generalize Eq.(20) to obtain the teleported (output) state
\begin{eqnarray}
\rho^{out}_{B_1B_2} 
& \equiv & \Lambda^{\chi_{A_1B_1} \otimes \chi_{A_2B_2},\ {\cal P}_1(m,\ n)}_{B_1B_2}(\rho^{in}_{B_1B_2}) \nonumber \\
& = & \sum^3_{i, j = 0}
({_{A_1B_1}}\langle\Psi^i_{Bell}|\chi_{A_1B_1}|\Psi^i_{Bell}\rangle_{A_1B_1} \times
 {_{A_2B_2}}\langle\Psi^j_{Bell}|\chi_{A_2B_2}|\Psi^j_{Bell}\rangle_{A_2B_2}) \nonumber \\
& & \times (\sigma^{i\oplus m}_{B_1} \otimes \sigma^{j\oplus n}_{B_2})\rho^{in}_{B_1B_2}
(\sigma^{i\oplus m}_{B_1} \otimes \sigma^{j\oplus n}_{B_2}),
\end{eqnarray}
where $m, n = 0, 1, 2, 3$, and \cite{Buzek}
\begin{eqnarray}
\rho^{in}_{B_1B_2} & = & \frac{1}{4}\{\sigma^0_{B_1} \otimes \sigma^0_{B_2} + \cos\mu
(\vec{r}_{B_1}\cdot\vec{\sigma}_{B_1} \otimes \sigma^0_{B_2} +
 \vec{r}_{B_2}\cdot\sigma^0_{B_1} \otimes \vec{\sigma}_{B_2}) \nonumber \\
& & + \sum^3_{i, j = 1}[r^i_{B_1}r^j_{B_2} 
+ \sin\mu\cos\nu(k^i_{B_1}k^j_{B_2} - l^i_{B_1}l^j_{B_2}) - \sin\mu\sin\nu(k^i_{B_1}l^j_{B_2} + l^i_{B_1}k^j_{B_2})]
\sigma^i_{B_1} \otimes \sigma^j_{B_2}\}, \nonumber \\
\end{eqnarray}
is the input state with $\mu,\ \vartheta_{B_1},\ \vartheta_{B_2} \in (0,\ \pi)$; $\nu,\ \varphi_{B_1},\ \varphi_{B_2} \in (0,\ 2\pi)$, and
\begin{eqnarray}
\vec{r}_{\alpha} & = & \left(\begin{array}{lll}
\sin\vartheta_{\alpha}\cos\varphi_{\alpha}, & \sin\vartheta_{\alpha}\sin\varphi_{\alpha},& \cos\vartheta_{\alpha}\end{array}\right),
\nonumber \\
\vec{k}_{\alpha} & = & \left(\begin{array}{lll}
\sin\varphi_{\alpha}, & -\cos\varphi_{\alpha}, & 0 \end{array}\right), \nonumber \\
\vec{l}_{\alpha} & = & \left(\begin{array}{lll}
\cos\vartheta_{\alpha}\cos\varphi_{\alpha}, & \cos\vartheta_{\alpha}\sin\varphi_{\alpha},& -\sin\vartheta_{\alpha}\end{array}\right).
\end{eqnarray}
Here, $\alpha = B_1$ or $B_2$.  It is easy to verify that $\rho^{in}_{B_1B_2}\cdot\rho^{in}_{B_1B_2} = \rho^{in}_{B_1B_2}$.  That is, $\rho^{in}_{B_1B_2}$ is an arbitrary unknown pure state of two qubits, which are in general entangled.

The entanglement teleportation fidelity, can be defined analogously to Eq.(21):
\begin{eqnarray}
\Phi[\Lambda^{\chi_{A_1B_1} \otimes \chi_{A_2B_2},\ {\cal P}_1(m,\ n)}_{B_1B_2}] & \equiv &
\frac{3}{64\pi^3}\int^{\pi}_0\int^{2\pi}_0\int^{\pi}_0\int^{2\pi}_0\int^{\pi}_0\int^{2\pi}_0
d\mu d\nu d\vartheta_{B_1} d\varphi_{B_1} d\vartheta_{B_2} d\varphi_{B_2} \nonumber \\
& & \times \cos^2\mu\sin\mu\sin\vartheta_{B_1}\sin\vartheta_{B_2}\ {\rm tr}[\rho^{in}_{B_1B_2}\rho^{out}_{B_1B_2}].
\end{eqnarray}
After some algebra, we obtain the maximal entanglement teleportation fidelity
\begin{equation}
\Phi_{\max}[\Lambda^{\chi_{A_1B_1} \otimes \chi_{A_2B_2},\ {\cal P}_1}_{B_1B_2}] = \left\{\begin{array}{lll}
\frac{1}{5}\left(1 + \frac{4}{Z^2}e^{2\beta J}\right) & {\rm for} & \sqrt{\eta^2 + \gamma^2} \leq 1 \\
\frac{1}{5}\left[1 + \frac{4}{Z^2}\left(\cosh\beta{\cal B} + \frac{\gamma J}{\cal B}\sinh\beta{\cal B}\right)^2\right] & {\rm for} & \sqrt{\eta^2 + \gamma^2} > 1
\end{array}\right.
\end{equation}
In terms of $\Phi_{\max}[\Lambda^{\chi_{A_1B_1} \otimes \chi_{A_2B_2},\ {\cal P}_1}_{B_1B_2}]$, for the thermal state Eq.(8) to be useful for entanglement teleportation, we demand that $\Phi_{\max}[\Lambda^{\chi_{A_1B_1} \otimes \chi_{A_2B_2},\ {\cal P}_1}_{B_1B_2}] > \frac{2}{5}$ \cite{Horodecki}.  This condition reduces to Eqs.(28) and (29) respectively.  Therefore, no new insights could be gained from the entanglement teleportation fidelity.

It is interesting to analyze the fidelity of the teleported state for some partially unknown pure input state of two qubits, which are in general entangled.  Consider, for instance,
\begin{equation}
|\Psi^{in}\rangle_{B_1B_2} = \cos\xi|00\rangle_{B_1B_2} + \sin\xi|11\rangle_{B_1B_2},
\end{equation}
where $0 \leq \xi \leq \pi$, with ${\cal C}[|\Psi^{in}\rangle_{B_1B_2}\langle\Psi^{in}|] = |\sin2\xi|$.  For $\sqrt{\eta^2 + \gamma^2} \leq 1$, we have the maximal output fidelity
\begin{eqnarray}
& & {_{B_1B_2}}\langle\Psi^{in}|
\Lambda^{\chi_{A_1B_1} \otimes \chi_{A_2B_2},\ {\cal P}_1(2,\ 2)}_{B_1B_2}(|\Psi^{in}\rangle_{B_1B_2}\langle\Psi^{in}|)
|\Psi^{in}\rangle_{B_1B_2} \nonumber \\
& = & \frac{1}{Z^2}\left[4\cosh^2\beta J + 2\left(1 + \frac{\gamma^2J^2}{{\cal B}^2}\right)\sinh^2\beta{\cal B}\sin^22\xi\right],
\end{eqnarray}
while for $\sqrt{\eta^2 + \gamma^2} > 1$,
\begin{eqnarray}
& & {_{B_1B_2}}\langle\Psi^{in}|
\Lambda^{\chi_{A_1B_1} \otimes \chi_{A_2B_2},\ {\cal P}_1(1,\ 1)}_{B_1B_2}(|\Psi^{in}\rangle_{B_1B_2}\langle\Psi^{in}|)
|\Psi^{in}\rangle_{B_1B_2} \nonumber \\
& = & \frac{1}{Z^2}\left\{
\left(e^{\beta J} + \cosh\beta{\cal B} + \frac{\gamma J}{\cal B}\sinh\beta{\cal B}\right)^2\right. \nonumber \\
& & + \left.\left[\left(e^{\beta J} + \cosh\beta{\cal B} - \frac{\gamma J}{\cal B}\sinh\beta{\cal B}\right)^2 - 2\sinh2\beta J - 4e^{\beta J}\cosh\beta{\cal B}\right]\sin^22\xi\right\}.
\end{eqnarray}
We note that, in contrast to the result in Ref.\cite{Lee}, the relation between the maximal output fidelity and the initial entanglement of the input state is not as straightforward.  When $\eta \leq \eta_{\rm critical} = \sqrt{1 - \gamma^2}$, the maximal output fidelity always increases monotonically as the initial entanglement increases for any $\beta, \gamma, \eta$, and $J$.  However, if $\eta > \eta_{\rm critical}$, it could be a monotonic increasing or decreasing function of the initial entanglement, depending on the choice of $\beta, \gamma, \eta$, and $J$.

\section{Conclusions}
In conclusion, we show that in an anisotropic two-qubit Heisenberg $XY$ chain, the nonzero thermal entanglement produced by adjusting the external magnetic field beyond some critical strength is a useful resource for teleportation via ${\cal P}_0$.  It would be interesting to determine if the same occurs in models such as those considered in Refs.\cite{Anteneodo} and \cite{Zhou}.  We also considered entanglement teleportation via two two-qubit Heisenberg $XY$ chains.  In particular, we show that for the partially unknown input state Eq.(36), the optimal output fidelity could be a monotonic increasing or decreasing function of the entanglement associated with the initial input state.

Y. Yeo thanks L. C. Kwek for useful discussions, and the Cambridge MIT Institute Limited for financial support.

\end{document}